\documentclass{aastex631}
\usepackage{graphicx} 
\usepackage{natbib}
\usepackage{tabularx}
\usepackage{mathtools}
\usepackage{wrapfig}
\usepackage{verbatim}
\usepackage{placeins}
\usepackage{subfigure}

\date{12 July 2024}

\begin{document}
\title{A New Parameterization for Finding Solutions for Microlensing Exoplanet Light Curves}

\author[0009-0006-1763-5936]{Kylie E. Hall}
\affiliation{Wellesley College Astronomy Department, 106 Central St., Wellesley, MA 02481, USA}
\email{kh1@wellesley.edu}

\author[0000-0001-9481-7123]{Jennifer C. Yee}
\affiliation{Center for Astrophysics $|$ Harvard \& Smithsonian, 60 Garden St.,Cambridge, MA 02138, USA}
\email{jyee@cfa.harvard.edu}

\author[0000-0002-4355-9838]{In-Gu Shin}
\affiliation{Center for Astrophysics $|$ Harvard \& Smithsonian, 60 Garden St.,Cambridge, MA 02138, USA}
\email{ingushin@gmail.com}

\author[0000-0003-0626-8465]{Hongjing Yang}
\affiliation{Department of Astronomy, Tsinghua University, Beijing 100084, China}

\author{Jiyuan Zhang}
\affiliation{Department of Astronomy, Tsinghua University, Beijing 100084, China}

\begin{abstract}
The gravitational microlensing method of discovering exoplanets and multi-star systems can produce degenerate solutions, some of which require in-depth analysis to uncover. We propose a new parameter space that can be used to sample potential solutions more efficiently and is more robust at finding all degenerate solutions for the ``central-resonant" caustic degeneracy. We identified two new parameters, $k$ and $h$, that can be sampled in place of the mass ratios and separations of the systems under analysis to identify degenerate solutions. The parameter $k$ is related to the size of the central caustic, $\Delta\xi_c$, while $h$ is related to the distance of a point along the $k$ contour from log($s$)=0, where $s$ is the projected planet-host separation. In this work, we present the characteristics of these parameters and the tests we conducted to prove their efficacy.  
\end{abstract}

\section{Introduction}
Gravitational microlensing is a method used to discover and characterize exoplanets \citep{liebes64,mao1991}. When a “lens” star passes in front of a “source” star relative to our line of sight, the gravity of the lens star bends the light of the source star, leading to a magnification pattern in the light curve of the source star \citep{einstein1936}. This pattern changes when the lens is composed of multiple bodies, for example a star hosting one or more planets. The mass ratio(s) of the star and planet(s), $q$, and the separation between the star and planet(s), $s$, can be derived from the magnification pattern in the light curve of such a microlensing event.

In general, the analysis of gravitational microlensing data is conducted by generating model light curves that assume particular characteristics (``parameters") of the lens and source systems and then using algorithms to fit those models to the observed data by tweaking the parameters of the model. 
Historically, the most commonly used algorithms to conduct fitting involve grid searches, Markov Chain Monte Carlo (MCMC) processes, or some combination of the two. More recently, the field has explored using other algorithms, including {\tt MultiNest} \citep{feroz2009,poleski2020} and differential evolution \citep{pyLIMA}, and also more complex approaches as in {\tt RTModel} \citep{RTModel}.

In the case of some events, multiple degenerate solutions can be found to fit observed data approximately equally as well. One of the earliest known degeneracies was the degeneracy between $s$ and $s^{-1}$ solutions, the so-called ``close-wide" degeneracy \citep{griest1998use,dominik1999binary}. Since then, many others have been encountered, some mathematical in nature and some ``accidental" \citep[i.e., due to poor light curve coverage from gaps in the observational data, cf.][]{skowron18accidental}. 
For example in \cite{bennett2008low}, sixteen different potential solution sets were identified for the same microlensing event (MOA-2007-BLG-192), which they attribute to four two-fold degeneracies including the ``close-wide" degeneracy and various parallax degeneracies \citep{Smith03,Gould04}. Because we cannot tell for sure which degenerate solution is the true solution, an algorithm might be missing the true solution if it misses a degenerate solution. Hence, it is important that all degenerate solutions are identified by the fitting algorithms used to analyze microlensing data. 

There are multiple examples of degenerate solutions being initially overlooked in past research. Section 3.1 of \citep{bachelet18} contains a review of several cases in which the standard grid-search procedure missed relevant solutions. More recently, \cite{yang2022kmt} identified another degeneracy that is easily missed in grid searches, that they dubbed the ``central-resonant" caustic degeneracy. Improving the performance of grid searches identifying the ``central-resonant" degeneracy was the primary motivation for this work.

Both microlensing events analyzed in \cite{yang2022kmt} were found to suffer from the ``central-resonant" caustic degeneracy, in which it could not be determined whether the caustic that caused the magnification pattern was a central or a resonant caustic. This results in multiple potential values for $s$, as well as other parameters. This degeneracy was not identified in the initial grid search for solutions, and required a denser grid search to be uncovered. A similar situation occurred in the analysis of two events by \cite{ryu2022mass}. All four events are drawn from a single year (2021), and \cite{yang2022kmt} notes that this relatively high frequency of occurrence indicates that the degeneracy may have been missed in previous events as well. In fact, once this degeneracy was recognized, re-analysis of OGLE-2016-BLG-1195 led to the discovery of new, previously--un-probed, solutions \citep{gould23ob1195}. Making grid searches more robust to the ``central-resonant" degeneracy is the primary motivation for this work.

Current grid search algorithms most often sample evenly in (log($s$), log($q$)) space. However, the local minima of microlensing event parameter spaces often follow a v-shaped pattern symmetrical about log($s$) $\simeq$ 0. This pattern roughly aligns with the boundaries in log($s$) and log($q$) separating different types of caustics. It might be more efficient to sample a higher density of test points within such a distribution instead of sampling evenly in log($s$) and log($q$). Sampling within this distribution might identify local minima that would otherwise require a higher-resolution grid search to unearth.

To construct a new grid that samples more densely from the region of interest, we found two new parameters related to $s$ and $q$ such that, when evenly spaced points in the space of these two new parameters are mapped to (log($s$), log($q$)) space, they follow a similar distribution pattern as the $\chi^2$ distribution seen in many microlensing events. In Section \ref{sec:kh}, we describe how we defined these parameters, why we expect them to be an improvement, and how we used them to construct a grid in log($s$) and log($q$) space. Then, in Section \ref{sec:tests}, we describe the tests we conducted to assess the efficacy of the new grid in practice. Finally, we conclude in Section \ref{sec:sum}.

{\section{New Parameters}
\label{sec:kh}}
\subsection{Defining k and h}

We defined the new parameters based on characteristics of the caustics associated with given values of $s$ and $q$, because these caustics change as $s$ and $q$ change. Our goal was to find a caustic characteristic that changes rapidly in the region of interest (near log($s$)=0 on either side). This characteristic would serve as the first parameter, $k$, and the distance along the contour lines of this parameter would serve as the second, $h$, so that points evenly spaced in these parameters will bunch up and become over-dense in that region when mapped back to log($s$) and log($q$).

We evaluated different caustic characteristics by plotting points in (log($s$), log($q$)) space with colors corresponding to the value of the caustic characteristic. The morphology of a given microlensing event is set by the magnification pattern on the source plane and that pattern can be summarized by the caustic(s). Therefore, it would be logical for some property of the caustic (or derivative) to be a more effective search variable, in part because it should capture correlations between $s$ and $q$. Then, if such a property could be identified, it might be possible to map it to some underlying mathematics. However, the goal of this work is primarily to identify a set of parameters that work empirically for the ``central-resonant" degeneracy, regardless of whether or not their theoretical origins can be identified.

Ultimately, we determined that the logarithm of the horizontal width of the central caustic, $\Delta\xi_c$, behaves in the desired manner \citep{chung2005properties}:
\begin{equation}
\label{eq:delta-xi}
    \Delta\xi_c = \frac{4q}{(s-s^{-1})^2}
\end{equation}
In some ways, this is surprising because Equation \ref{eq:delta-xi} was calculated in the limit that $q\ll1$ and only for central caustics. Nevertheless, we are applying this equation over all of ($s$, $q$) space because it matches the desired behavior even if it does not necessarily describe the caustic accurately over the whole space.

Previously, \citet{dong2009} used a similar parameter, $\Delta\eta_c$ (the extent of the central caustic in the y-direction), to conduct a grid search for solutions to MOA-2007-BLG-400. They argued that this parameter would be useful in cases for which it could be estimated directly from the light curve. Here, we use the measurement along the x-axis, which has a simpler analytic form. Similar to the observation by \citet{dong2009} about $\Delta\eta_c$, $\Delta\xi_c$ is linear with $\log s$ in many regimes\footnote{In fact, $\Delta\eta_c \propto \Delta\xi_c$ \citep[see Eq. 11 and 12 of ][]{chung2005properties}.}.

\begin{figure}[t!]
    \centering
    \includegraphics[width=\textwidth]{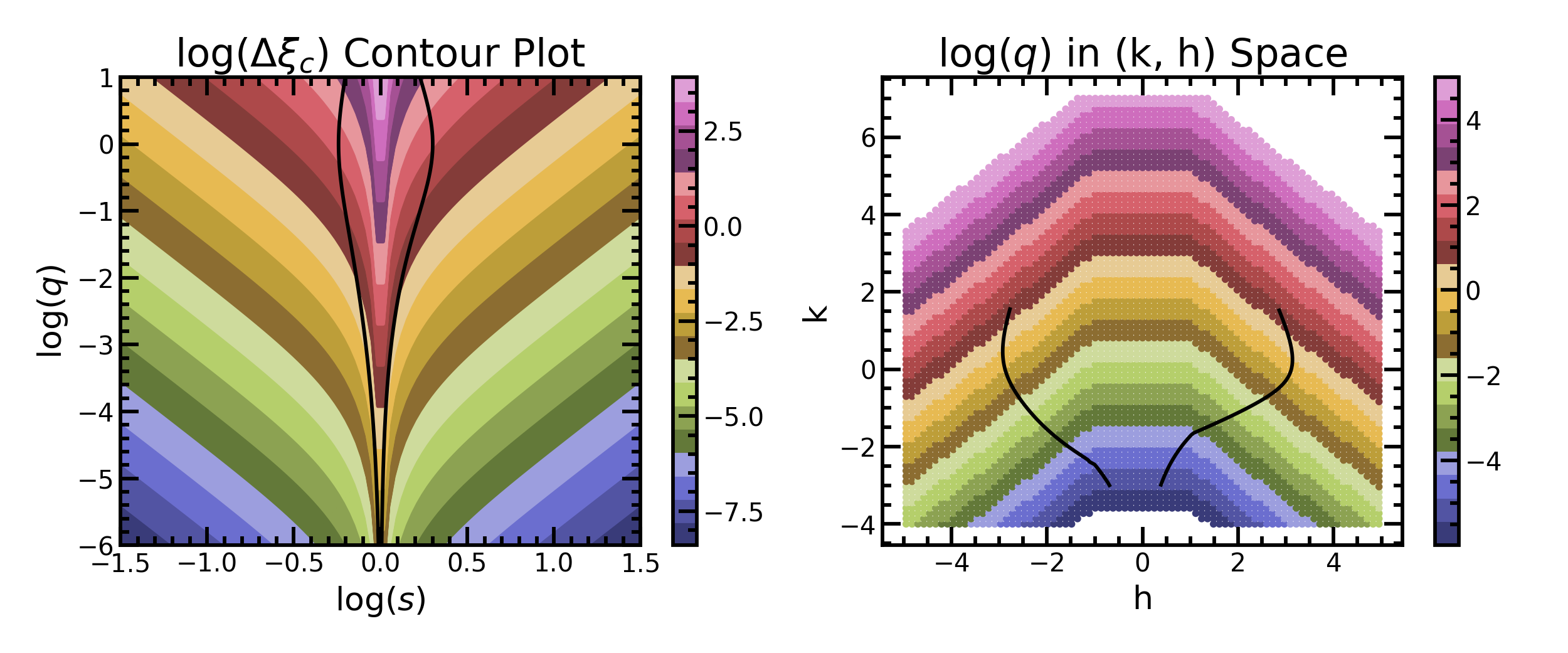}
    \caption{
    (Left) Contour lines for constant values of $\log(\Delta\xi_c)$ in (log($s$), log($q$)) space. Black lines represent the boundaries between different caustic types \cite{dominik1999binary}, with close caustic geometries on the left, wide on the right, and resonant in between the lines. These contour lines become denser in the region where microlensing solutions are more likely to be located, making $\log(\Delta\xi_c)$ a good parameter to sample evenly when searching for microlensing solutions. The flattening of the contours near $\log (s) \sim 0$ is an artifact of the plotting algorithm; in reality the contours asymptote as they approach $\log (s) = 0$, see Equation \ref{eq:delta-xi}. (Right) An evenly-spaced grid in ($k$, $h$) space color-coded by corresponding values of log($q$). Black lines represent the caustic boundaries translated into ($k$, $h$) space.}
    \label{fig:logq_k-h}
\end{figure}

As shown in Figure \ref{fig:logq_k-h}, contour lines of $\log_{10}(\Delta\xi_c)$ in (log($s$), log($q$)) space 
become denser in the target region of (log($s$), log($q$)) space (i.e., as $\log s$ approaches the resonant caustic regime) as desired. 
However, as can be seen from Equation \ref{eq:delta-xi}, the contours also asymptote as they approach log($s$)=0, which would result in infinitely dense contours as log($s$) approaches 0. Thus, to ensure that $k$ is defined at all values of log($s$), we defined both $k$ and $h$ as piece-wise functions with different definitions near log($s$)=0. We define a transition point at $\log (s_{\rm ref}) = 0.03$. This value was chosen empirically to ensure that a grid search with a similar number of points as in \citet{yang2022kmt} would have multiple grid points covering the resonant caustic minima.

So, for $|\log(s)| \geq \log (s_{\rm ref})$:
\begin{equation}
    k(s, q) \equiv \log_{10}(\Delta\xi_c) = \log \left( \frac{4q}{(s-s^{-1})^2} \right).
    \label{eq:k}
\end{equation}

For $|\log(s)| < \log (s_{\rm ref})$, we hold $k$ constant over changing log($s$):
\begin{equation}
    k(s, q) = \log\left(\frac{4 q}{(s_{\rm ref} - s_{\rm ref}^{-1})^2}\right) = \log(q) + 2.3;
\end{equation}
i.e., for each point (log($s_i$), log($q_i$)) within this range, we set $k$ equal to the value of $\log(\Delta\xi_c)$ at the point $( \pm \log(s_{\rm ref}), \log(q_i) )$.

For the parameter $h$, we want a definition that generally reflects a distance from $\log (s) = 0$. For $|\log (s)| \gg 0$, the $k$ contours become approximately straight lines in ($\log(s)$, $\log(q)$) space so they may be described as a simple magnitude equation, but this breaks down as $|\log (s)| \rightarrow 0$ because the slope of $k$ changes dramatically. Of course, this slope change is the behavior that gets us the higher density of contours for $|\log s| \rightarrow 0$. So, for $|\log (s)| < \log (s_{\rm ref})$, we define $h$ to maintain 
the density of points just outside this region. 

Hence,
for $\log(s) \geq \log (s_{\rm ref})$:
\begin{equation}
    h = \sqrt{[\log(s) - \log(s_{\rm ref})]^2 + 
              [\log(q)-\log(q_{\rm ref})]^2} + C.
\end{equation}
The value of $q_{\rm ref}$ is defined as follows: for a given point ($s$, $q$), the value of $k$ can be determined from Equation \ref{eq:k}. Then, $q_{\rm ref}$ is the value of $q$ that satisfies Equation \ref{eq:k} for that value of $k$ and $s=s_{\rm ref}$.

For $-\log (s_{\rm ref}) < \log(s) < \log (s_{\rm ref}):$
\begin{equation}
    h = m \log(s).
\end{equation}

Finally, for $\log(s) \le -\log (s_{\rm ref})$:
\begin{equation}
    h = -\sqrt{[\log(s) - \log(s_{\rm ref})]^2 + 
               [\log(q)-\log(q_{\rm ref})]^2} - C,
\end{equation}
where $C\equiv m \log(s_{\rm ref})$ such that the contours match up at $\log(s_{\rm ref})$. We choose $m = 35$ because it maintains the approximate $h$ spacing near the transition point.

A grid of points in ($k$, $h$) space, color-coded by values of log($q$), can be found in Figure \ref{fig:logq_k-h}.

\subsection{Evaluating the New Grid}

\cite{yang2022kmt} ultimately needed to perform three grid searches of increasing densities and increasingly narrow parameter ranges in order to uncover all 4 degenerate solutions for KMT-2021-BLG-0171 and all 6 degenerate solutions for KMT-2021-BLG-1689, because their initial grid was not dense enough in the regions where the solutions were located to find all solutions. Therefore, to estimate the efficacy of the new grid, we compared the new grid to the initial (log($s$), log($q$)) grid employed by \cite{yang2022kmt} in their analysis of events KMT-2021-BLG-0171 and KMT-2021-BLG-1689. 

\begin{figure}[t]
    \centering
    \includegraphics[width=\columnwidth]{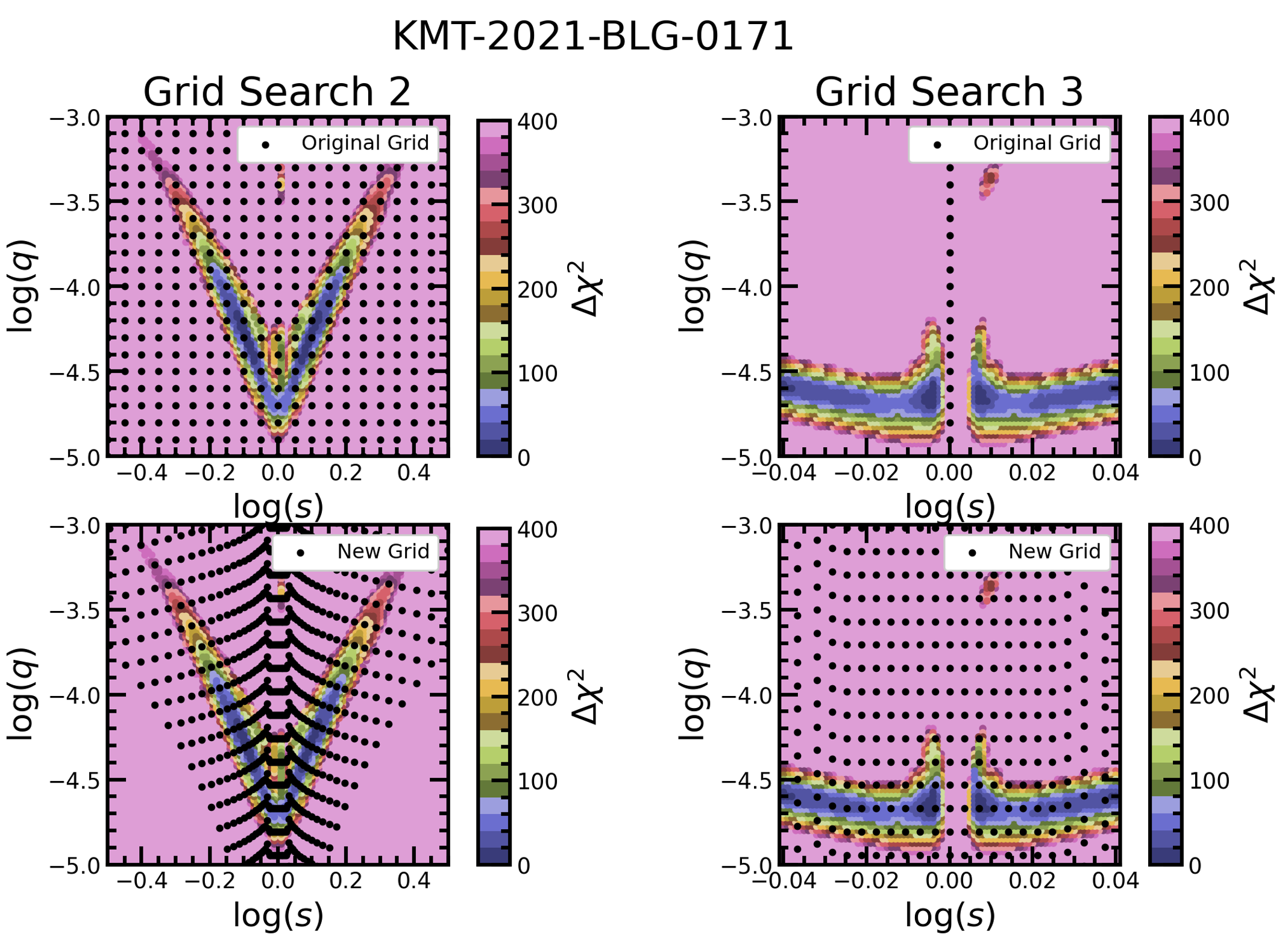}
    \caption{
    A comparison of the behaviors of the first grid of \cite{yang2022kmt} (top) vs. the new grid (bottom) plotted over the $\chi^2$ distribution found by the second (left) and third (right) grid searches of \cite{yang2022kmt} for event KMT-2021-BLG-0171. Although the two grids have about the same number of points within -1.5 $<\log(s)<$ 1.5 and -6 $<\log(q)<$ 1, the ($k$, $h$) grid has many more points in the region where the solutions for this event fall, meaning the ($k$, $h$) grid should be more efficient at picking up on these solutions.}
    \label{fig:compare-0171}
\end{figure}

\begin{figure}[t]
    \centering
    \includegraphics[width=\columnwidth]{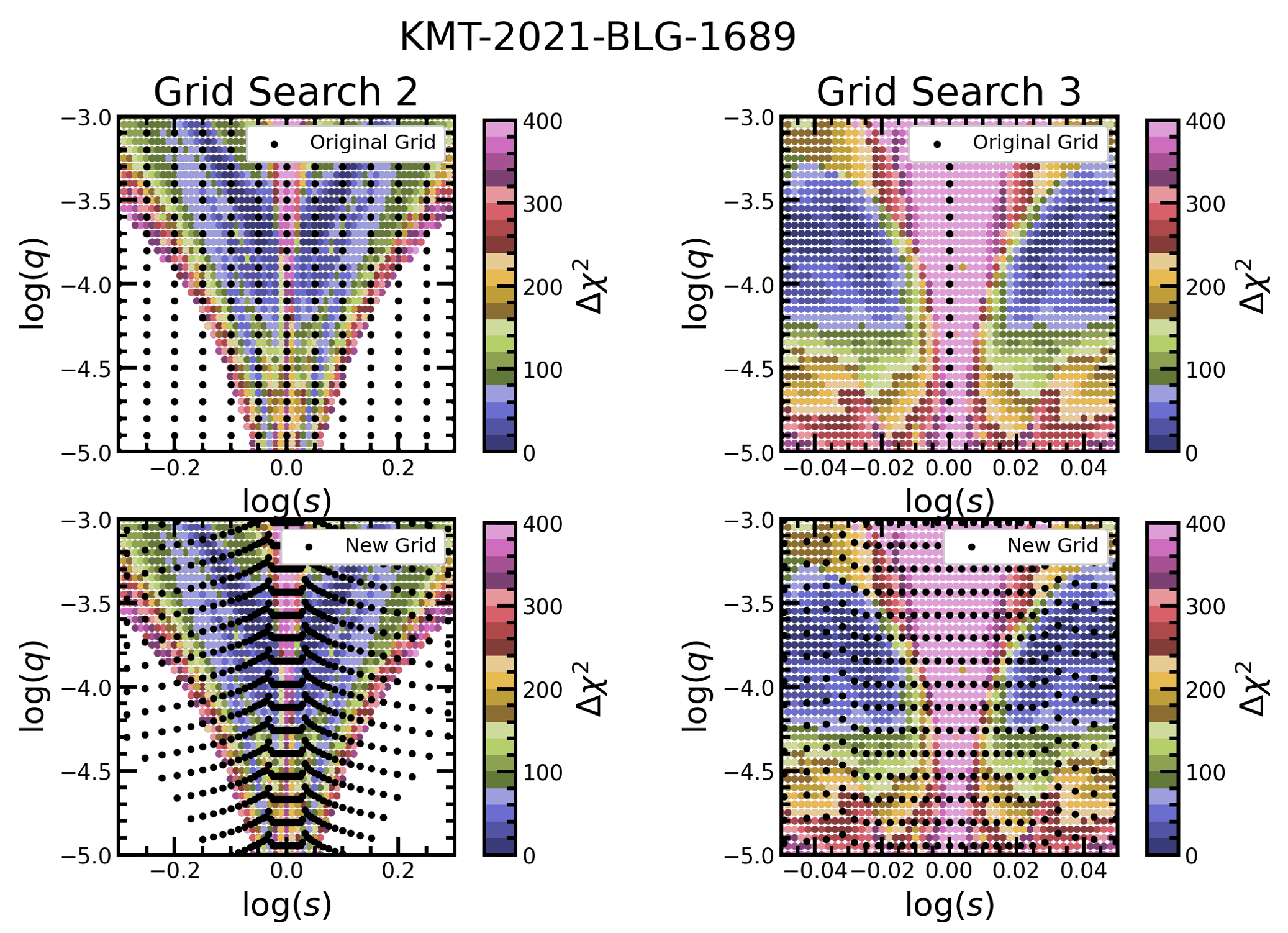}
    \caption{
    As in Figure \ref{fig:compare-0171} for event KMT-2021-BLG-1689.}
    \label{fig:compare-1689}
\end{figure}

Figures \ref{fig:compare-0171} and \ref{fig:compare-1689} show both the new grid points and the initial grid search points used by \cite{yang2022kmt} overlaid on the $\chi^2$ distributions found by the densest grid searches performed by \cite{yang2022kmt} for events KMT-2021-BLG-0171 (for which two degenerate pairs of solutions were found) and KMT-2021-BLG-1689 (for which three degenerate pairs of solutions were found). Each of the grids have approximately the same number of total points covering the range between -1.5 $<\log(s)<$ 1.5 and -6 $<\log(q)<$ 1 (the new grid has slightly fewer). In principle, solutions with $\log q > 0$ are symmetric with those for $\log q < 0$, but in order to ensure continuity in parameter searches (e.g. if $k$ and $h$ were free parameters), we calculate the grid including $\log q$ up to 1.  As can be seen from these plots, the new grid contains many more points within the local minima than the initial grid from \cite{yang2022kmt}, which suggests that the new grid might be better at identifying these minima. 

\begin{figure}
    \centering
    \begin{subfigure}{}
        \includegraphics[width=\textwidth]{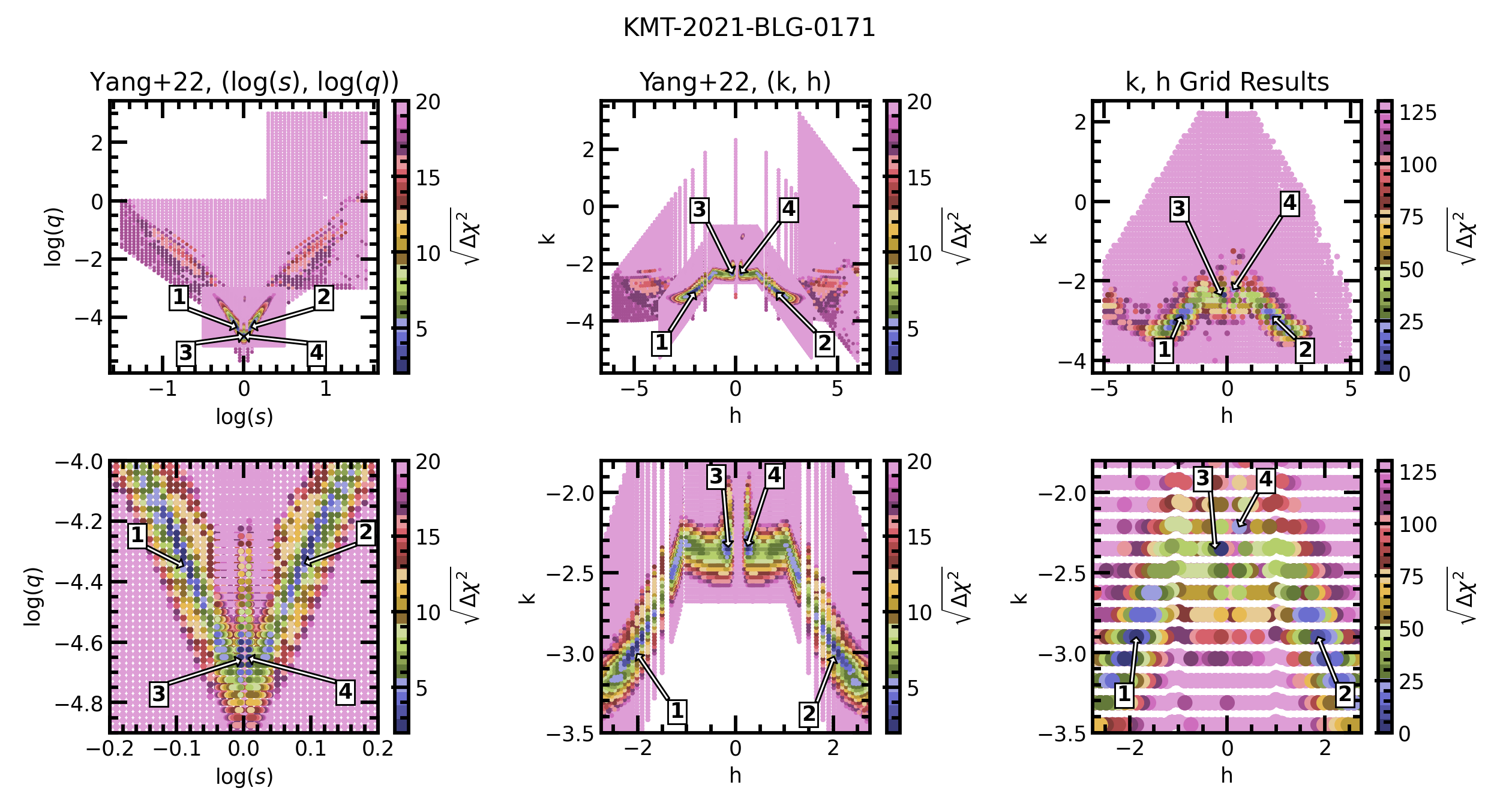}
    \end{subfigure}
    \begin{subfigure}{}
        \includegraphics[width=\textwidth]{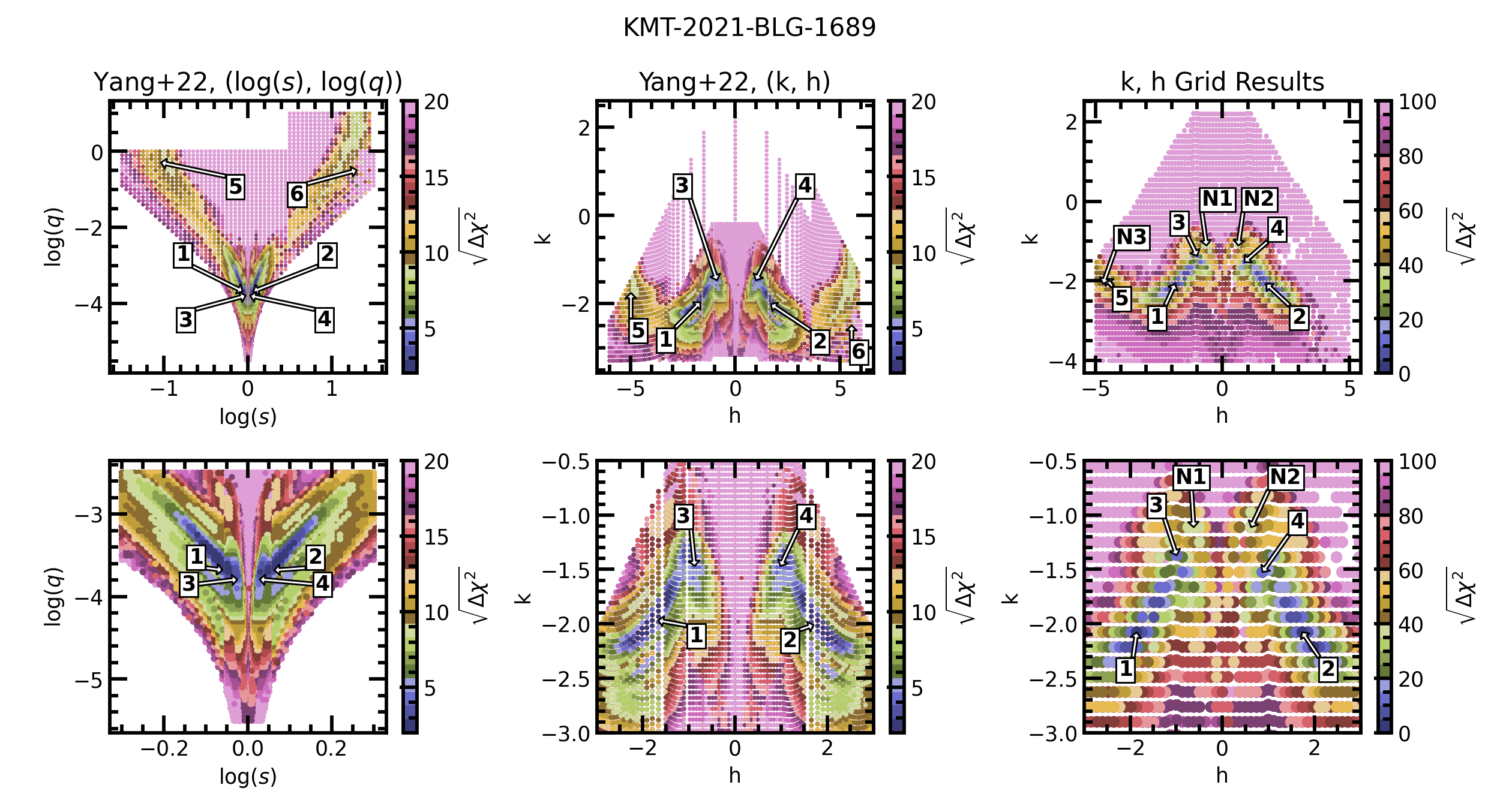}
    \end{subfigure}
    \caption{The leftmost and middle plots displayed are the $\Delta\chi^2$ distribution found by \cite{yang2022kmt} for events KMT-2021-BLG-0171 (top two rows) and KMT-2021-BLG-1689 (bottom two rows) in three increasingly dense and narrow grid searches, plotted in both (log(s), log(q)) space (left) and ($k, h$) space (middle). The top plots in each pair of rows shows the full extent of the grid search whereas the bottom plots show a zoom. Solutions are labeled by number (see Tables \ref{tab:0171_sols} and \ref{tab:1689_sols}). In (log($s$), log($q$)) space, many of the solutions are too close together to be resolved. These solutions are better separated in ($k, h$) space. The rightmost plots contain grid search results using the ($k, h$) grid for these events. All four solutions found by \cite{yang2022kmt} with three grid searches for event KMT-2021-BLG-0171 were recovered using only one grid search in ($k, h$) space. Five of the six solutions found by \cite{yang2022kmt} with three grid searches for event KMT-2021-BLG-1689 were recovered using only one grid search in ($k, h$) space. Three new minima were also found for this event.}
    \label{fig:0171-1689-plots}
\end{figure}

We also investigated the efficacy of the new parameters by calculating ($k$, $h$) values for the points in the grid searches conducted by \cite{yang2022kmt} and re-plotting the $\chi^2$ distribution in ($k$, $h$) space. These plots can be found in Figure \ref{fig:0171-1689-plots}. As can be seen in this figure, the different minima are well-separated in ($k$, $h$) space, suggesting that a reasonably dense grid of points evenly spaced in $k$ and $h$ would be able to resolve all of the minima. 

Ideally, this grid would be effective at finding solutions for a variety of microlensing events, not just planetary microlensing events. Thus, to test whether binary star system solutions would be resolvable in ($k$, $h$) space, we repeated this interpolation process for the $\chi^2$ distributions found using a (log($s$), log($q$)) grid for four additional events, all with binary-star-system lens solutions.
These events are KMT-2016-BLG-0020, KMT-2016-BLG-0157, KMT-2016-BLG-0199, KMT-2016-BLG-2542. They were chosen from events modeled in the AnomalyFinder \citep{Zang21AF1} search for planets in the 2016 prime fields \citep{Shin23AF9} and selected for having a variety of morphologies for the $\chi^2$ surface in (log($s$), log($q$))-space.

\begin{figure*}
    \centering
    \includegraphics{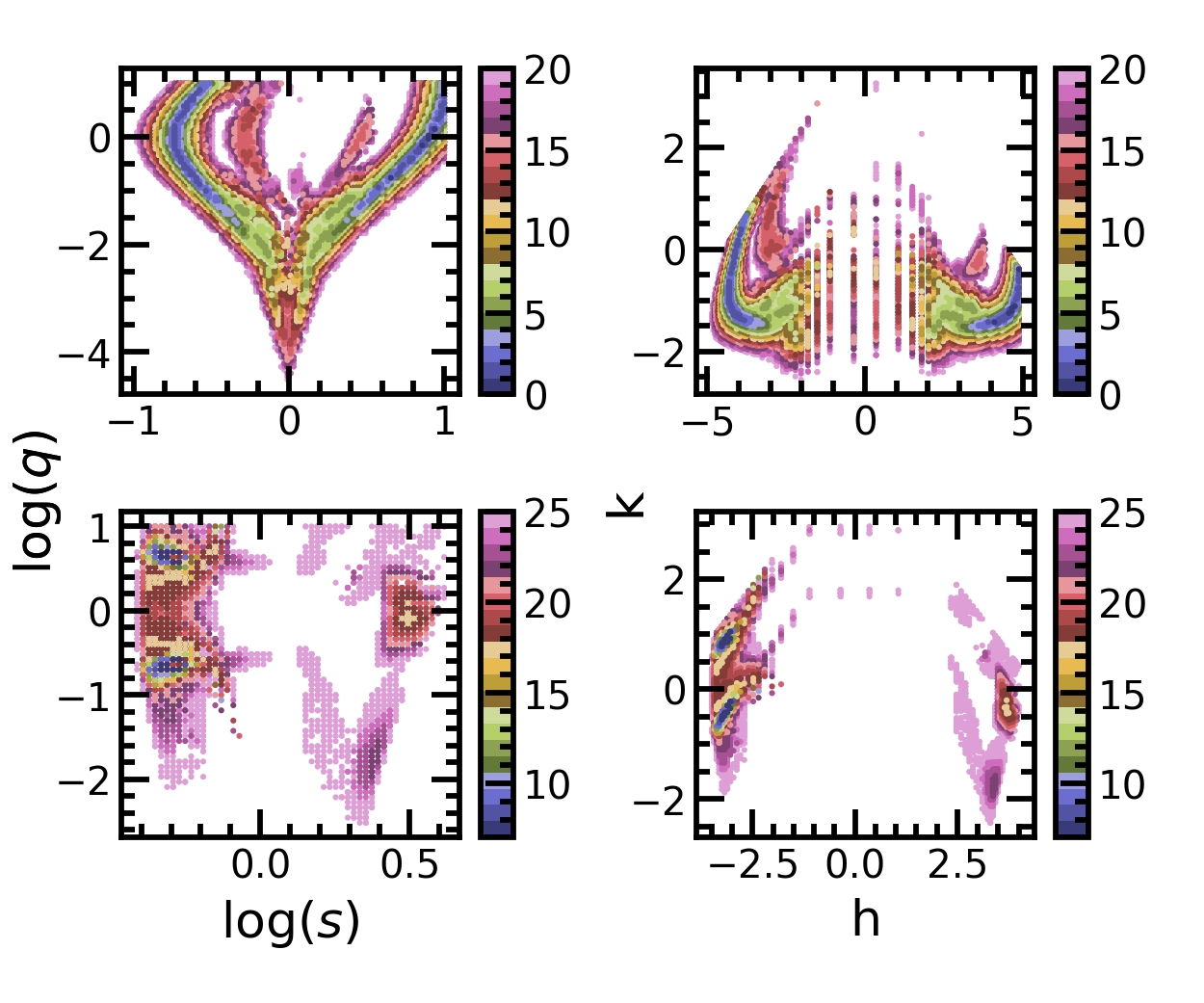}
    \caption{ 
    The $\chi^2$ distribution for events KMT-2016-BLG-2542 (top) and KMT-2016-BLG-0199 (bottom), both with binary lens solutions originally found using a grid evenly spaced in (log($s$), log($q$)) space (left) and then re-plotted in ($k, h$) space (right). Colors indicate values of $\sqrt{\Delta\chi^2}$. In all cases, solutions that were resolvable in (log($s$), log($q$)) space are still resolvable in ($k, h$) space, meaning a grid of points evenly spaced in ($k, h$) could be used effectively to analyze events with binary system solutions.}
    \label{fig:ex-grids_1}
\end{figure*}

\begin{figure}
    \centering
    \includegraphics{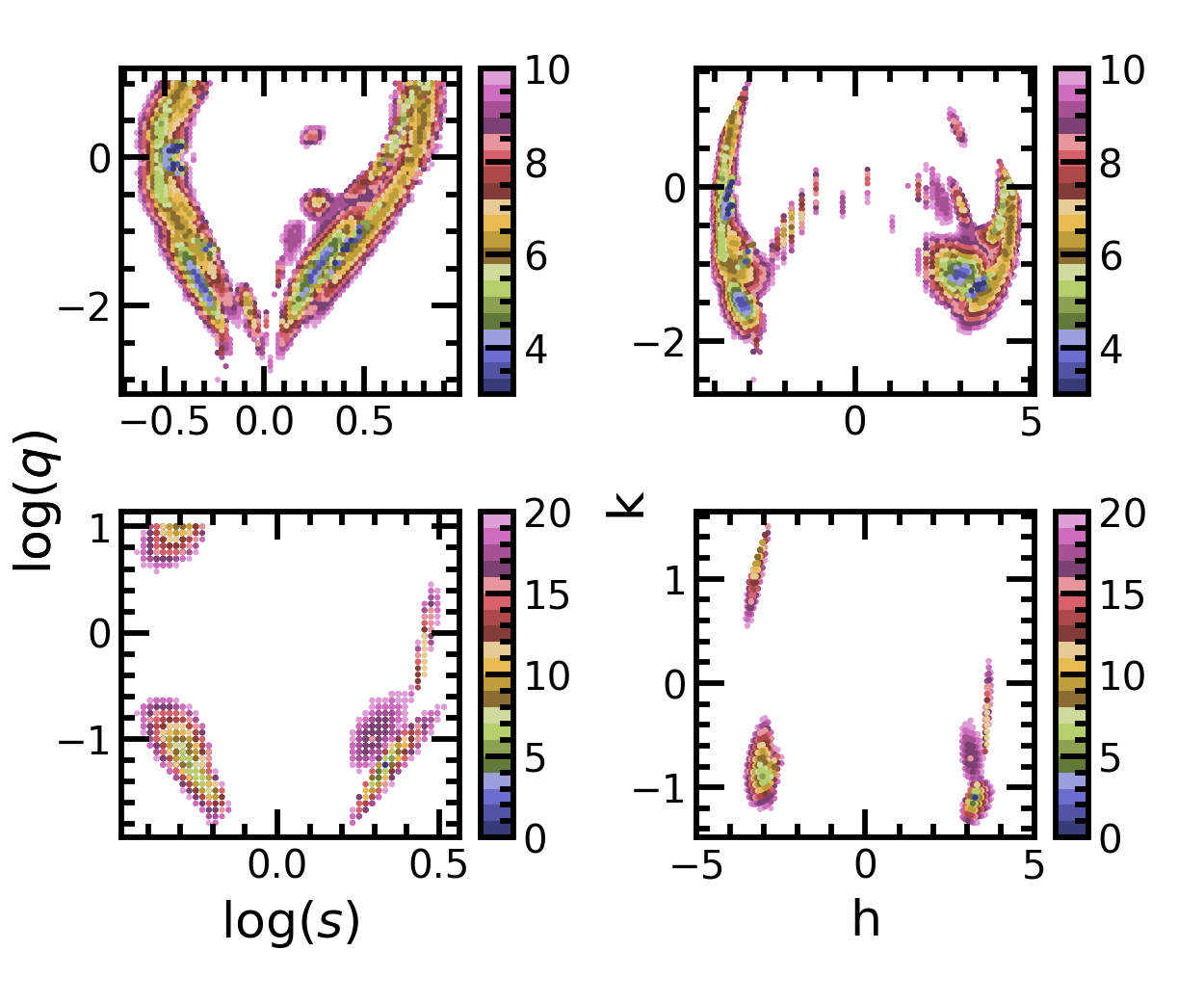}
    \caption{As in Figure \ref{fig:ex-grids_1} for events KMT-2016-BLG-0020 (top) and KMT-2016-BLG-0157 (bottom).}
    \label{fig:ex-grids_2}
\end{figure}

The (log($s$), log($q$)) grids as well as their transforms into ($k$, $h$)-space can be found in Figures \ref{fig:ex-grids_1} and \ref{fig:ex-grids_2}. As can be seen in these figures, all of the minima resolvable in (log($s$), log($q$)) space are also resolvable in ($k$, $h$) space. Thus, the ($k$, $h$) grid should also be effective at analyzing non-planetary events. 

For the present work, we will focus verifying that $k$ and $h$ are better parameters for resolving minima for the central-resonant degeneracy in a few specific cases.

{\section{Tests}
\label{sec:tests}}

In order to prove the efficacy of the new grid, we conducted a grid search/MCMC analysis of three different previously-analyzed microlensing events as case tests: KMT-2021-BLG-0171 and KMT-2021-BLG-1689 \citep{yang2022kmt} and MOA-2007-BLG-192 \citep{bennett2008low}. These events each suffer from multiple degeneracies. 

For each search, we used {\tt VBBL} \citep{bozza2018vbbinarylensing} with {\tt MulensModel} \citep{poleski2019modeling} to generate the light curves and {\tt emcee} \citep{emcee2013} to refine the parameters, keeping $s$ and $q$ fixed. We allowed the source size, $\rho$, to be a free parameter, but to simplify the calculations did not include limb-darkening in our models.  The version of {\tt MulensModel} we used ({\tt v2.11}) does not allow for $q > 1$, so we exclude any ($k$, $h$) values that produce $q>1$ from our grid search. Prior to the search, we also renormalized the error bars of each dataset by a constant factor so that the $\chi^2/\mathrm{d.o.f.} = 1$ relative to a point lens model.

The following section will describe each event in more detail and present the results of our analysis.

\subsection{KMT-2021-BLG-0171}
The solutions previously found for this event are listed in Table \ref{tab:0171_sols}. This event suffers from a degeneracy between the (1, 2) and (3, 4) solutions. Both 1 and 2 predict the same values for source size $\rho$ and $q$, and both 3 and 4 predict the same values for $\rho$ and $q$, but the values predicted by (1, 2) and (3, 4) differ from each other for both parameters. Additionally, (1, 2) predict a larger absolute value for $\log(s)$ than (3, 4) (known as the ``central-resonant" degeneracy). Within each pair, the ``close-wide" degeneracy is also present. This is a common microlensing degeneracy where two solutions exist that are identical except that the value for $s$ has approximately undergone a $s$ $\xleftrightarrow{}$ $s^{-1}$ transformation. See \cite{griest1998use} and \cite{dominik1999binary} for further discussion of this degeneracy, and see \cite{yang2022kmt} for more information on this event and its solutions.

The ($k$, $h$) grid was successful in identifying all four solutions identified by \cite{yang2022kmt} in a single grid search, in contrast to the three grid searches required in \cite{yang2022kmt}. See Figure \ref{fig:0171-1689-plots} for a plot of the $\chi^2$ distribution from our grid search and Table \ref{tab:0171_sols} for a comparison of the solutions found by the ($k$, $h$) grid and the solutions found by \cite{yang2022kmt}.

\subsection{KMT-2021-BLG-1689}
This event suffers from a degeneracy between the (1, 2) and (3, 4) solutions. The two pairs predict different values for $\rho$ and $q$. This event also suffers from the central-resonant degeneracy between solution pairs (1, 2) and (3, 4), and solution pairs (1, 2), (3, 4), and (5, 6) each suffer from the close-wide degeneracy. See Table \ref{tab:1689_sols} and \cite{yang2022kmt} for more information on this event and its solutions.

The ($k$, $h$) grid was successful in finding solutions 1 through 5 in a single grid search, compared to the three grid searches used by \cite{yang2022kmt}. It did not find a minimum near the location of solution 6. 
However, that solution has the secondary as the center of magnification, i.e., the source passes by the ``planetary" caustic rather than the central caustic. Hence, finding that solution would generally require changing the origin of the coordinate system to the secondary or searching for solutions with $q > 1$.
See Figure \ref{fig:0171-1689-plots} for a plot of the $\chi^2$ distribution from this grid search and Table \ref{tab:1689_sols} for a comparison of the solutions found by the ($k$, $h$) grid and the solutions found by \cite{yang2022kmt}.

\begin{figure}
    \centering
    \includegraphics[width=140mm]{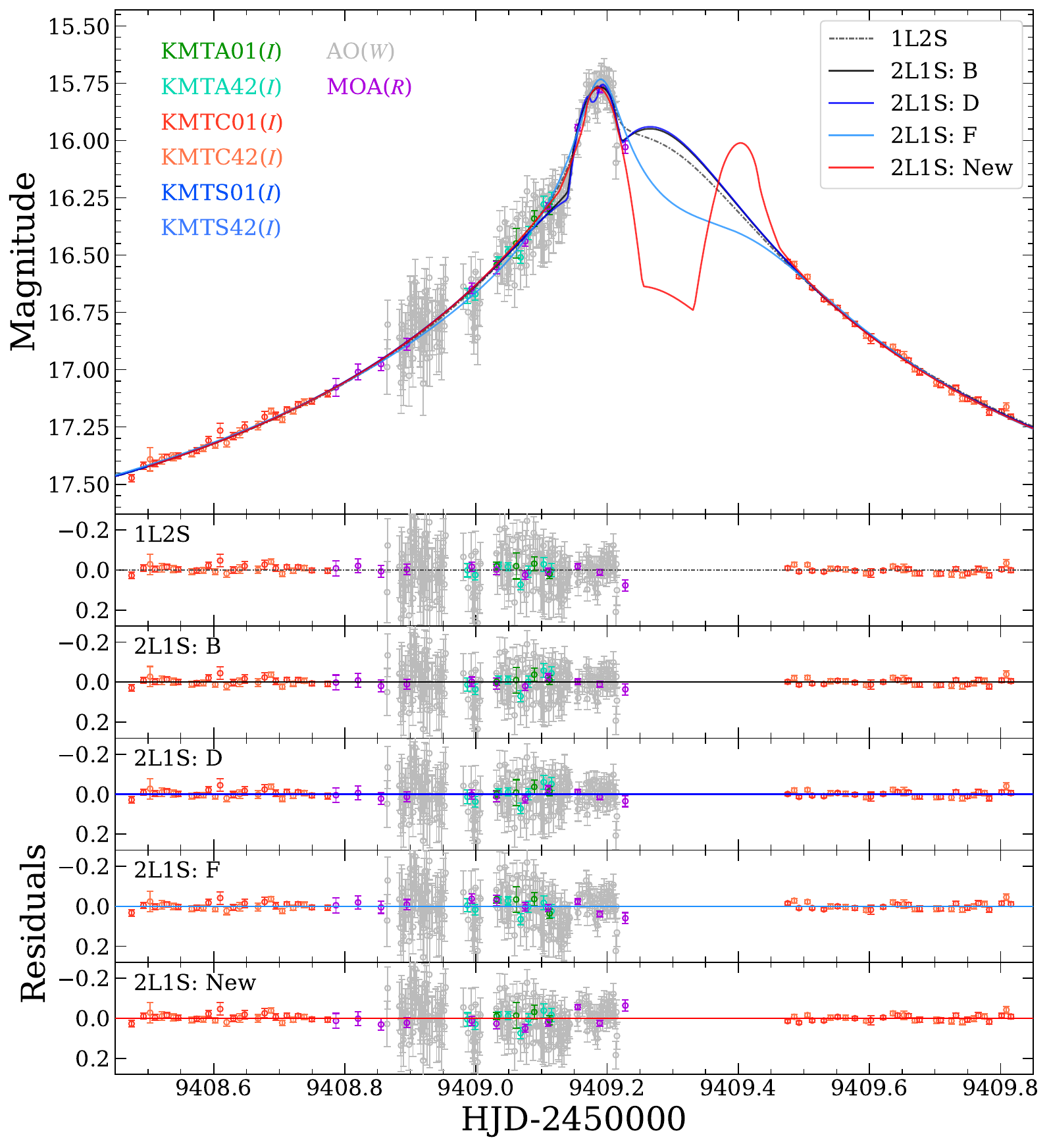}
    \caption{Solutions for event KMT-2021-BLG-1689, with the light curve corresponding to the new minimum found by the ($k, h$) grid shown in red. This minimum was ultimately rejected as a solution due to its high $\Delta\chi^2$ value, although it is a better fit than Solution F shown here. Note that Solutions B, D, and F are referred to as Solutions 2, 4, and 6 throughout this paper.}
    \label{fig:1689sols-with-new-min}
\end{figure}

The ($k$, $h$) grid also appeared to have three new minima, not previously identified by \cite{yang2022kmt}, two of which represent a potential new solution suffering from the close-wide degeneracy. The parameters of these minima from the grid search, as well as their $\Delta\chi^2$ relative to the best-fit solution from our grid search are given in Table \ref{tab:1689_sols}. In addition, 
we ran a refined MCMC analysis on one of these minima and found the set of parameters with the best $\chi^2$ value listed in Table \ref{tab:1689_sols}, with the corresponding light curve shown in Figure \ref{fig:1689sols-with-new-min}. This results in a $\Delta\chi^2=61.4$ relative to the best-fit solutions in \cite{yang2022kmt}. So, 
these minima were investigated as potential new solutions but were ultimately rejected due to their high $\Delta\chi^2$ values ($\Delta\chi^2> 50$). However, they are a better fit to the data than the binary solutions (5, 6) investigated in \cite{yang2022kmt}. Thus, the fact that they were discovered by the ($k$, $h$) grid suggests that this grid has the potential to uncover solutions that would be missed by the traditional (log($s$), log($q$)) grid.

\subsection{MOA-2007-BLG-192}

This event is challenging to analyze in part due to the lack of data coverage of the event. In addition, two four-fold degeneracies contribute to the 16 solutions reported for this event by \cite{bennett2008low}. The most relevant degeneracies for this work are those which produce variation in log($s$) and log($q$) amongst the solutions; one such degeneracy is due to an undersampling of data which led to an inability to distinguish whether a caustic crossing vs. cusp approaching model is superior, and the other is the close-wide degeneracy. We list a few of these solutions in Table \ref{tab:192_sols}. These solutions reflect the four basic morphologies of the solutions reported in Table 3 of \cite{bennett2008low}. See \cite{bennett2008low} for more information on this event and its solutions. 

\begin{figure}
    \centering
    \includegraphics[width=\textwidth]{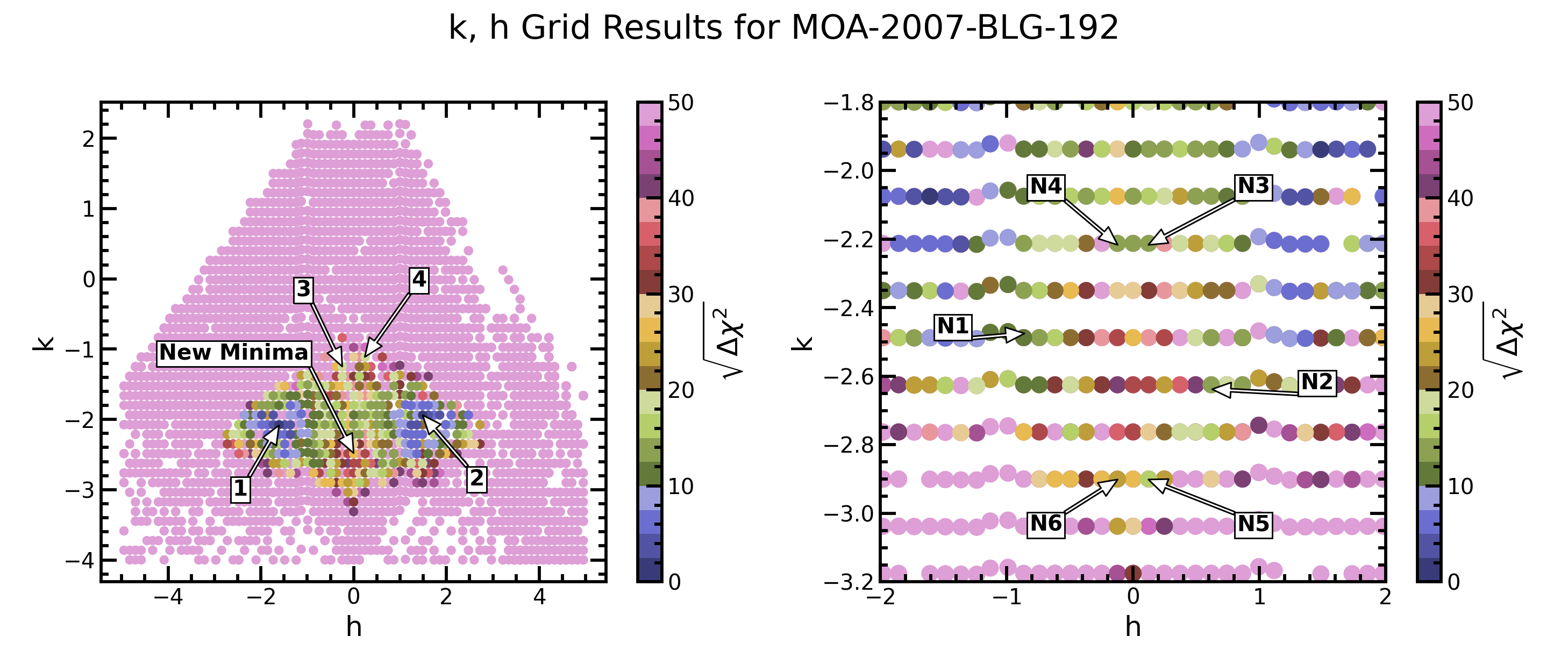}
    \caption{Grid search results using the ($k, h$) grid for event MOA-2007-BLG-192. All four planetary solutions with distinct $s$ and $q$ values found by \cite{bennett2008low} were recovered using only one grid search (left panel). In addition, our grid search reveals a complex $\chi^2$ surface near $h\sim 0$. We identified a number of additional, potential solutions in this region (right panel).}
    \label{fig:192-results}
\end{figure}

Figure \ref{fig:192-results} shows a plot of the $\chi^2$ distribution from this grid search and Table \ref{tab:192_sols} for a comparison of the solutions found by the ($k, h$) grid and the solutions found by \cite{bennett2008low}.
The ($k, h$) grid was successful in finding equivalents to all 4 solutions from \cite{bennett2008low} given in Table \ref{tab:192_sols}. We were also able to find equivalents to  the other three minima that can be identified from Figure 7 of \cite{bennett2008low},  but whose solutions are not reported in their table.

\begin{figure}
	\includegraphics[width=0.5\textwidth]{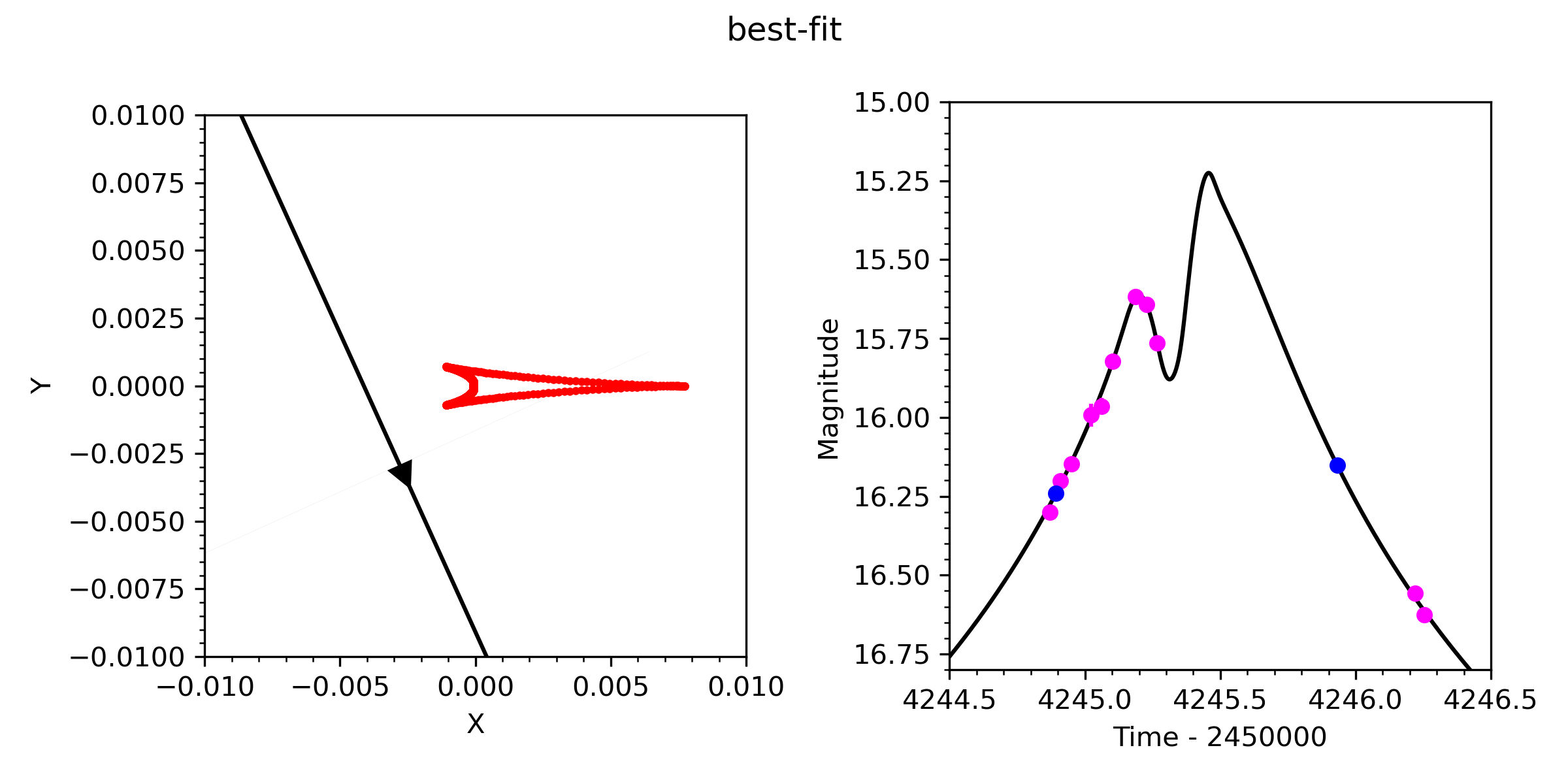}
	\includegraphics[width=0.5\textwidth]{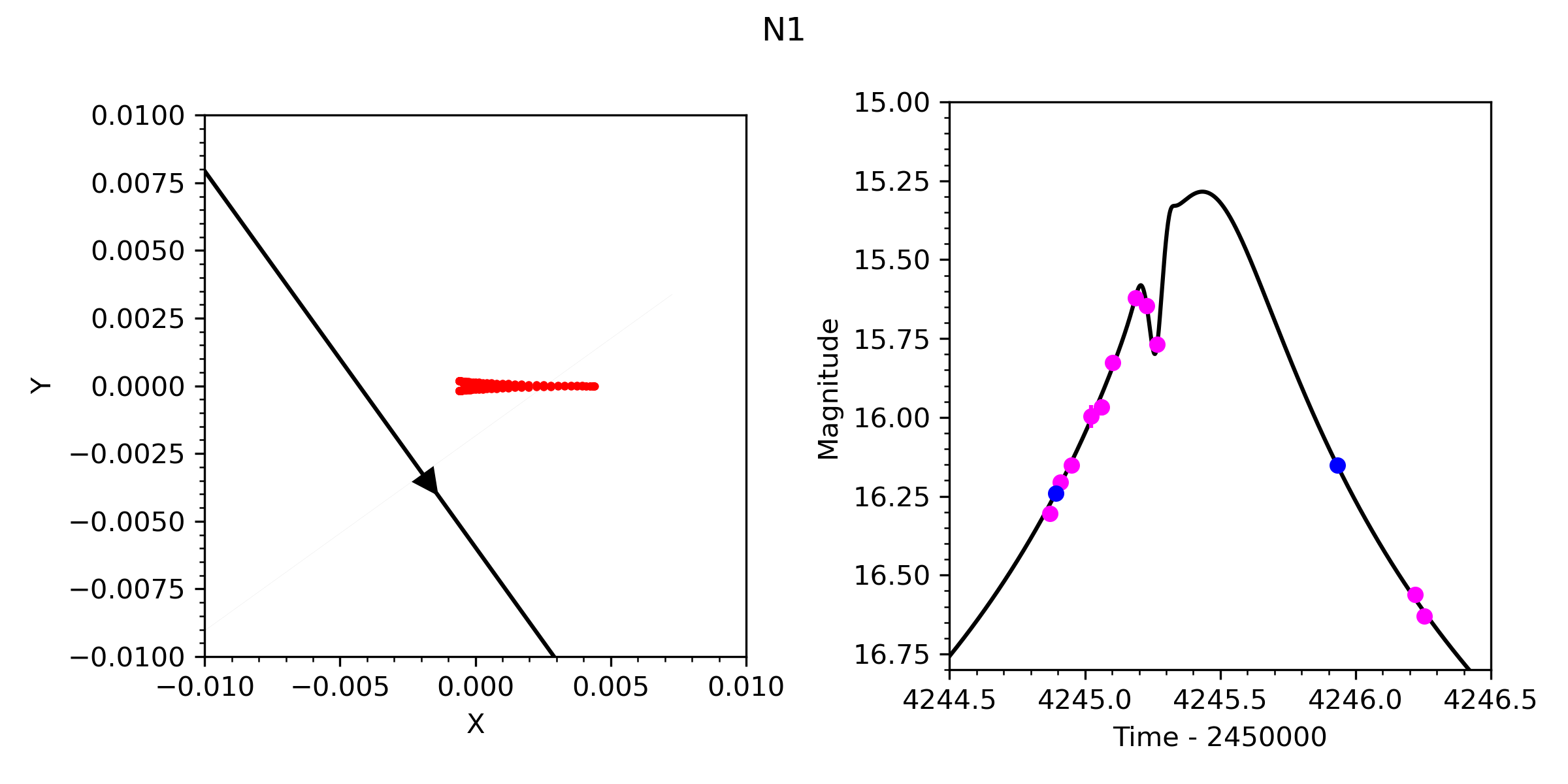}
	\includegraphics[width=0.5\textwidth]{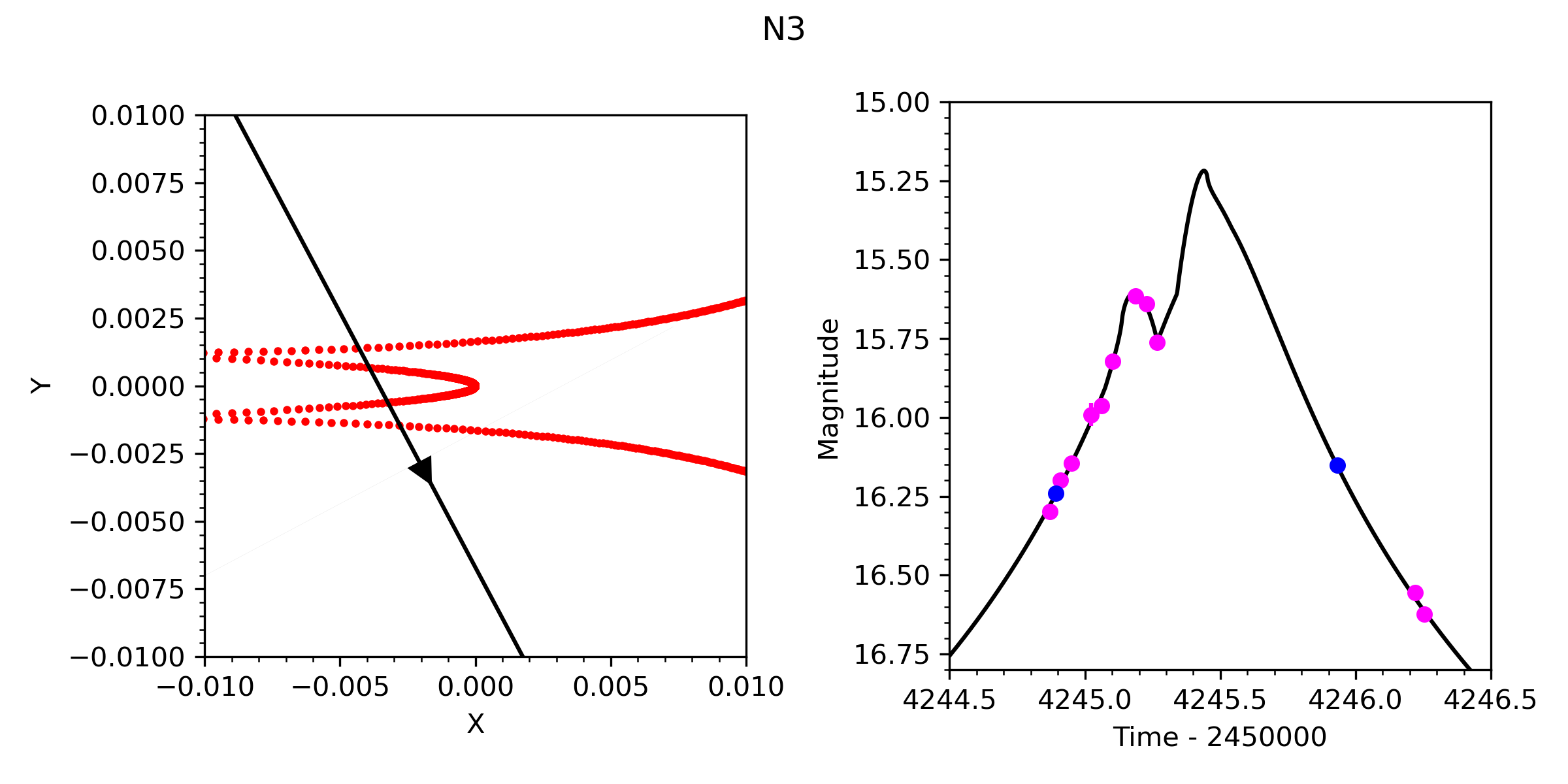}
	\includegraphics[width=0.5\textwidth]{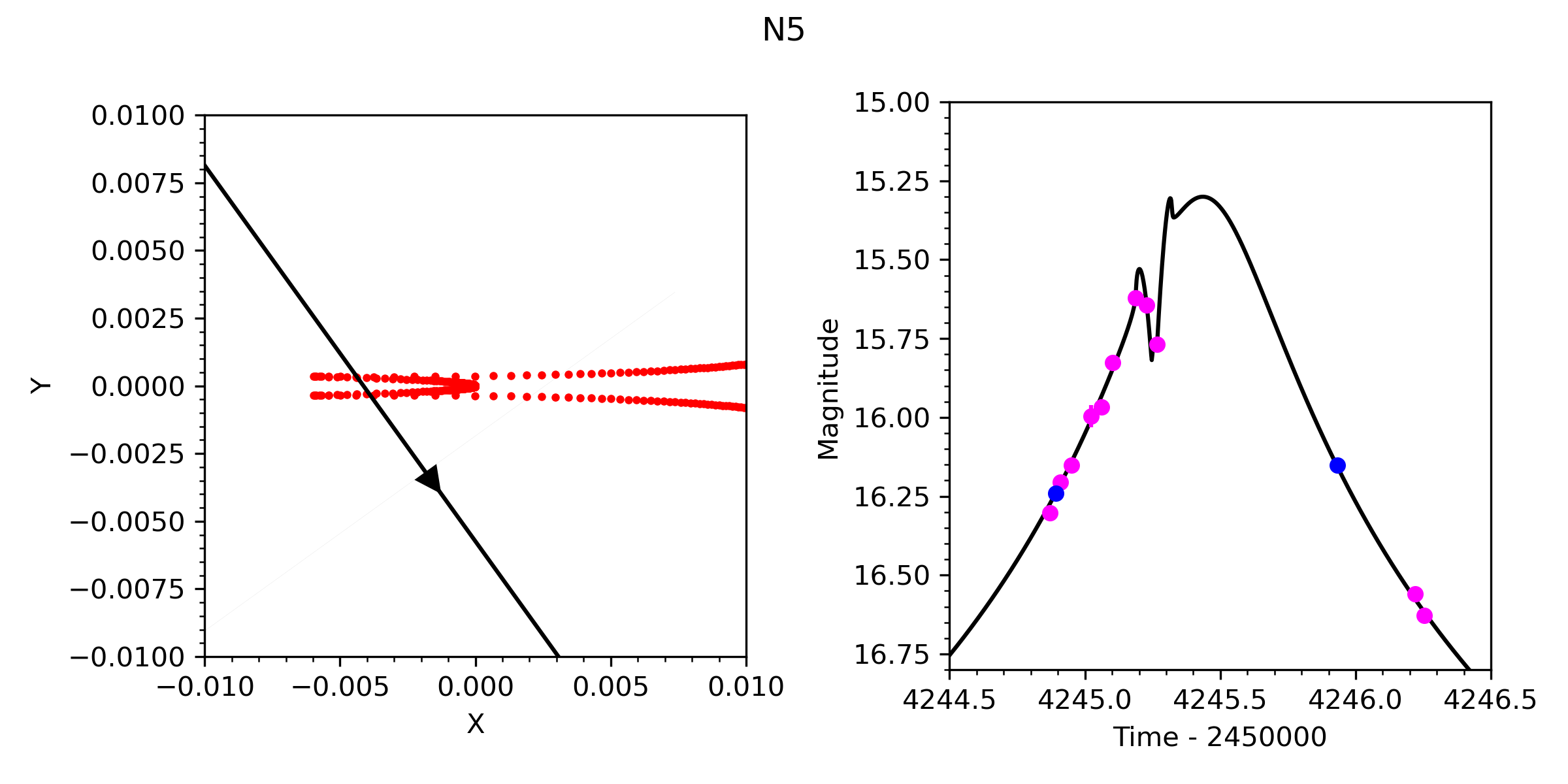}
	\caption{Our grid search for MOA-2007-BLG-192 identified several potential new minima with somewhat different morphologies from the best-fit (upper left). {\it Left panels}: Caustic structures (red) and source trajectories (black). {\it Right panels}: model light curves (black) with data from MOA (magenta) and OGLE (blue). \label{fig:mb192_lcs}}
\end{figure}

In addition, our $k, h$ grid shows significant structure for $|h| \lesssim 1$. In this region, we identified three families of solutions comprising six solutions, with each pair connected by the close-wide degeneracy. Figure \ref{fig:mb192_lcs} compares the light curves and source trajectories of the best solution in each pair with our best-fit solution. The main difference in the solutions is the location of the ``dip" in the light curve with respect to the final MOA observation from the night of HJD$^{\prime}=4245$. In the primary solution, the ``dip" occurs after the final MOA observation. However, in solutions N1/N2 and N5/N6, the ``dip" occurs between the final and penultimate points. In N3/N4, the final point occurs at the exact minimum of the ``dip".

Understanding the exact relationship between these various minima, e.g., by optimizing the individual minima, is beyond the scope of this work.\footnote{Among other things, a recent paper \citep{Terry24} has shown that the MOA data from the original paper, which we used in our analysis, have systematics that affect the modeling.} However, it is clear that the transformation from $(s, q)$- to $(k, h)$- space will facilitate such investigations.

{\section{Summary}
\label{sec:sum}}
Our goal in this work was to increase the analytic sensitivity of gravitational microlensing parameter searches by creating a grid search method more sensitive to the degenerate solutions observed in \cite{yang2022kmt}. We accomplished this goal by sampling points to be tested as solutions in a new parameter space that requires a lower density of points to be able to identify solutions compared to parameter spaces more commonly used. We defined these parameters by looking for caustic characteristics that change rapidly in the regions of (log($s$), log($q$)) space where degenerate solutions where more likely to be located, so that grid points sampled evenly in these parameters would be more likely to pick up on solutions. We ultimately defined these parameters based on the width of the central caustic of a given event. 

We tested a grid sampled evenly in these new parameters, which we name $k$ and $h$, on three events with degenerate solutions and found that the new grid could more efficiently identify most if not all of the solutions identified by previous analyses. In addition, for two of the events, we identified new potential solutions. We thus present the $k$ and $h$ parameters as a potential tool to be used in the analysis of gravitational microlensing events in the future. Given that this tool is more sensitive to the central-resonant degeneracy, it provides the opportunity for increased automation and efficiency in the analysis of gravitational microlensing events.
This will be especially important for microlensing searches with the {\em Roman Space Telescope}.

\section*{Acknowledgments}
The authors would like to thank the anonymous referee who made a number of useful comments that improved the paper.
J.C.Y. and I.-G.S. acknowledge support from U.S. NSF Grant No. AST-2108414. 
This research has made use of publicly available data 
(https://kmtnet.kasi.re.kr/ulens/) from the KMTNet system
operated by the Korea Astronomy and Space Science Institute
(KASI) at three host sites of CTIO in Chile, SAAO in South
Africa, and SSO in Australia. Data transfer from the host site to
KASI was supported by the Korea Research Environment
Open NETwork (KREONET).
The computations in this paper were conducted on the Smithsonian High Performance Cluster (SI/HPC), Smithsonian Institution. https://doi.org/10.25572/SIHPC.
This research has made use of the NASA Exoplanet Archive, which is operated by the California Institute of Technology, under contract with the National Aeronautics and Space Administration under the Exoplanet Exploration Program.

\FloatBarrier
\newpage

\begin{deluxetable}{llrrrrrrr}
\label{tab:0171_sols}
\setlength{\tabcolsep}{12pt}
    \centering
    \tablecaption{KMT-2021-BLG-0171 Solutions}
\tablehead{
        \colhead{Grid} & \colhead{Solution} & \colhead{t$_0$ (HJD)} & \colhead{u$_0$} & \colhead{t$_E$ (d)} & \colhead{$\rho$ (10$^{-3}$)} & \colhead{$\alpha$ (deg)} & \colhead{$s$} & \colhead{$q$ (10$^{-5}$)}
 }
\startdata
         Yang+22 & 1 & 2459326.2338 & 0.00564 & 41.57 & 1.50 & 237.6 & 0.813 & 4.28 \\
         & & $\pm$ 0.0003 & $\pm$ 0.00005 & $\pm$ 0.32 & $\pm$ 0.015 & $\pm$ 0.7 & $\pm$ 0.032 & $\pm$ 0.80 \\
         $k$, $h$ & & 2459326.2336 &  0.00568 & 41.42 & 1.71 &  236.9 & 0.841 & 3.79 \\
         \\
         Yang+22 & 2 & 2459326.2338 & 0.00564 & 41.56 & 1.51 & 237.7 & 1.232 & 4.17 \\
         & & $\pm$ 0.0003 & $\pm$ 0.00005 & $\pm$ 0.32 & $\pm$ 0.015 & $\pm$ 0.7 & $\pm$ 0.051 & $\pm$ 0.82 \\
         $k$, $h$ & & 2459326.2331 &  0.00567 & 41.42 & 1.76 &  236.7 & 1.190 & 3.84 \\
         \\
         Yang+22 & 3 & 2459326.2338 & 0.00565 & 41.57 & 1.62 & 239.1 & 0.9905 & 2.19 \\
         & & $\pm$ 0.0003 & $\pm$ 0.00005 & $\pm$ 0.32 & $\pm$ 0.007 & $\pm$ 0.4 & $\pm$ 0.0009 & $\pm$ 0.14 \\
         $k$, $h$ & & 2459326.2333 &  0.00568 & 41.45 & 1.81 &  237.3 & 0.984 & 2.13 \\
         \\
         Yang+22 & 4 & 2459326.2338 & 0.00565 & 41.55 & 1.62 & 239.2 & 1.0161 & 2.22 \\
         & & $\pm$ 0.0003 & $\pm$ 0.00005 & $\pm$ 0.31 & $\pm$ 0.007 & $\pm$ 0.4 & $\pm$ 0.0009 & $\pm$ 0.15 \\
         $k$, $h$ & & 2459326.2335 &  0.00575 & 40.97 & 1.85 &  238.1 & 1.016 & 2.93 \\
         \enddata
    \tablecomments{For each solution, values from \cite{yang2022kmt} are listed first, then uncertainty values from \cite{yang2022kmt}, then values found using the ($k$, $h$) grid. Solutions from \cite{yang2022kmt} were refined using an MCMC algorithm allowing $s$, $q$, and $\rho$ to be free parameters, while the ($k$, $h$) grid solutions were unrefined. This and the lack of limb-darkening may account for some of the discrepancy between the solutions from \cite{yang2022kmt} and the ($k$, $h$) grid solutions.}
\end{deluxetable}

\begin{deluxetable}{llrrrrrrr}
\label{tab:1689_sols}
\setlength{\tabcolsep}{12pt}
    \centering
    \tablecaption{KMT-2021-BLG-1689 Solutions}
\tablehead{
        \colhead{Grid} & \colhead{Solution} & \colhead{t$_0$ (HJD)} & \colhead{u$_0$} & \colhead{t$_E$ (d)} & \colhead{$\rho$ (10$^{-3}$)} & \colhead{$\alpha$ (deg)} & \colhead{$s$} & \colhead{$q$ (10$^{-4}$)}
 }
\startdata
    \multicolumn{9}{l}{Solutions Reported in Yang+22:}\\
         Yang+22 & 1 & 2459409.2510 & 0.00600 & 22.56 & 1.44 & 242.3 & 0.870 & 2.10 \\
         & & $\pm$ 0.0011 & $\pm$ 0.00028 & $\pm$ 0.84 & $\pm$ 0.08 & $\pm$ 0.6 & $\pm$ 0.025 & $\pm$ 0.39 \\
         $k, h$ & & 2459409.2529 &  0.00618 & 22.34 & 1.43 &  242.2 & 0.841 & 2.53 \\
                  \\
         Yang+22 & 2 & 2459409.2509 & 0.00601 & 22.51 & 1.44 & 242.3 & 1.157 & 2.09 \\
         & & $\pm$ 0.0011 & $\pm$ 0.00026 & $\pm$ 0.79 & $\pm$ 0.08 & $\pm$ 0.6 & $\pm$ 0.032 & $\pm$ 0.37 \\
         $k, h$ &
& 2459409.2525 &  0.00593 & 22.76 & 1.31 &  241.7 & 1.163 & 1.93 \\
          \\
         Yang+22 & 3 & 2459409.2509 & 0.00590 & 22.61 & 0.70 & 242.1 & 0.944 & 1.62 \\
         & & $\pm$ 0.0012 & $\pm$ 0.00027 & $\pm$ 0.85 & $\pm$ 0.08 & $\pm$ 0.6 & $\pm$ 0.004 & $\pm$ 0.17 \\
         $k, h$ && 2459409.2527 &  0.00613 & 22.16 & 0.74 &  242.3 & 0.937 & 2.04 \\
         \\
         Yang+22 & 4 & 2459409.2510 & 0.00587 & 22.78 & 0.68 & 242.2 & 1.067 & 1.62 \\
         & & $\pm$ 0.0011 & $\pm$ 0.00027 & $\pm$ 0.81 & $\pm$ 0.08 & $\pm$ 0.5 & $\pm$ 0.005 & $\pm$ 0.18 \\
         $k, h$ && 2459409.2526 &  0.00603 & 22.33 & 0.64 &  241.5 & 1.059 & 1.43 \\
            \\
         Yang+22 & 5 & 2459409.2403 & 0.00663 & 22.92 & $<$ 1.2 & 340.9 & 0.092 & 5079 \\
         & & $\pm$ 0.0012 & $\pm$ 0.00032 & $\pm$ 0.88 & -- & $\pm$ 1.0 & $\pm$ 0.006 & $\pm$ 2232 \\
         $k, h$  && 2459409.2522 &  0.00752 & 21.45 & 0.18 &  151.2 & 0.139 & 1436.59 \\
		         \\
         Yang+22 & 6 & 2459409.2394 & 0.00327 & 46.14 & $<$ 0.8 & 160.8 & 19.97 & 3186 \\
         & & $\pm$ 0.0009 & $\pm$ 0.00060 & $\pm$ 8.48 & -- & $\pm$ 0.5 & $\pm$ 1.25 & $\pm$ 1979 \\
         $k, h$ & & Not Found &  &  &  &  &  &  \\
         \hline
         	\multicolumn{9}{l}{Additional Minima from $k, h$ search:}\\
	$\Delta\chi^2$ &\\
 54.6 & N1 
& 2459409.2509 &  0.00689 & 22.43 & 1.92 &  100.9 & 0.960 & 3.69 \\
	& Optimized & 
		2459409.2493 & 0.00694 & 23.04 &    1.91 & 104.1 & 0.951 &  4.57 \\
   && $\pm$0.0015 & $\pm$0.00035 &  $\pm$0.84 &    $\pm$0.09 & $\pm$0.021 & $\pm$0.2 &  $\pm$0.36 \\
   \\
 58.1 & N2 
& 2459409.2511 &  0.00654 & 23.60 & 1.86 &  100.6 & 1.042 & 3.69 \\
 74.5 & N3 
& 2459409.2486 &  0.00669 & 22.73 & 0.52 &  246.4 & 0.123 & 1359.44 \\
\enddata
    \tablecomments{As in Table \ref{tab:0171_sols} for event KMT-2021-BLG-1689. The $\Delta\chi^2$ of the new solutions are given relative to the best-fit from the $k, h$ grid search.}
\end{deluxetable}

\begin{deluxetable}{llrrrrrrr}
\label{tab:192_sols}
\setlength{\tabcolsep}{12pt}
    \centering
        \tablecaption{MOA-2007-BLG-192 Solutions}
\tablehead{
        \colhead{Grid} & \colhead{Solution} & \colhead{t$_0$ (HJD)} & \colhead{u$_0$} & \colhead{t$_E$ (d)} & \colhead{$\rho$ (10$^{-4}$)} & \colhead{$\alpha$ (deg)} & \colhead{$s$} & \colhead{$q$ (10$^{-4}$)}
 }
\startdata
	\multicolumn{9}{l}{Solutions Reported in Table 3 of Bennett+08:}\\
         Bennett+08 & 1 & 2454245.453 & -0.00364 & 75.0 & 8.93 & 113.6 & 0.881 & 1.5 \\
         $k, h$ & & 2454245.448 & -0.00375 & 72.4 & 7.69 &  294.4 & 0.878 & 1.4 \\
         \\
         Bennett+08 & 2 & 2454245.453 & -0.00360 & 74.5 & 8.59 & 115.8 & 1.120 & 1.2 \\
         $k, h$ & & 2454245.448 & -0.00398 & 70.4 & 8.85 &  293.2 & 1.121 & 1.5 \\
         \\
         Bennett+08 & 3 & 2454245.462 & -0.00433 & 75.1 & 15.6 & 101.1 & 0.985 & 2.1 \\
         $k, h$ & & 2454245.441 & -0.00384 & 96.1 & 12.22 &  269.9 & 0.984 & 2.7 \\
         \\
         Bennett+08 & 4 & 2454245.458 & -0.00420 & 74.9 & 15.2 & 103.8 & 1.007 & 1.6 \\
         $k, h$ && 2454245.431 & -0.00464 & 85.1 & 13.95 &  265.0 & 1.016 & 3.7 \\
         \hline
	\multicolumn{9}{l}{$k, h$ Equivalents to Additional Minima from Figure 7 of  Bennett+08:}\\
	& & 2454245.443 & -0.00401 & 69.9 & 12.67 &  293.4 & 1.000 & 0.6 \\
	&& 2454245.415 & -0.00522 & 74.1 & 7.84 &  260.0 & 0.740 & 10.8 \\
	&& 2454245.388 & -0.00669 & 69.6 & 6.77 &  255.6 & 1.416 & 14.5 \\
	\hline
	\multicolumn{9}{l}{Additional Minima from $k, h$ search:}\\
	$\Delta\chi^2$ &\\
  5.1 & N1 
& 2454245.440 & -0.00348 & 71.6 & 3.26 &  305.7 & 0.945 & 0.16 \\
  6.5 & N2 
& 2454245.438 & -0.00344 & 72.3 & 1.93 &  306.5 & 1.042 & 0.11 \\
  7.6 & N3 
& 2454245.442 & -0.00314 & 83.8 & 10.21 &  298.0 & 1.008 & 0.29 \\
  8.1 & N4 
& 2454245.440 & -0.00374 & 72.4 & 11.71 &  296.7 & 0.992 & 0.29 \\
 12.1 & N5 
& 2454245.438 & -0.00335 & 75.2 & 4.10 &  305.8 & 1.008 & 0.06 \\
 20.8 & N6 
& 2454245.435 & -0.00418 & 62.5 & 4.64 &  304.6 & 0.992 & 0.06 \\
\enddata
 \tablecomments{
 As in Table \ref{tab:1689_sols} for event MOA-2007-BLG-192.
  In addition, due to a difference in the choice of coordinate system, there is a 180 degree offset between $\alpha$ from \citet{bennett2008low} and our values.}
\end{deluxetable}


\FloatBarrier

\end{document}